\documentclass[a4paper,fleqn,usenatbib]{mnras}
\usepackage{newtxtext,newtxmath}
\usepackage[T1]{fontenc}
\usepackage{ae,aecompl}
\usepackage{graphicx}
\usepackage{amssymb}
\usepackage{amsmath}
\usepackage[english]{babel}

\usepackage{array}
\newcolumntype{L}{>{$}l<{$}}

\title{Constraining the star formation rate with the extragalactic background light}
\author[A. A. Korochkin, G. I. Rubtsov]{A. A. Korochkin$^{1,2}$\thanks{E-mail:aa.korochkin@physics.msu.ru}, G. I. Rubtsov$^{1}$ \\
$^{1}$ Institute for Nuclear Research of the Russian Academy of Sciences, 60th October Anniversary st. 7a, 117312, Moscow, Russia\\
$^{2}$ Faculty of Physics, M~V~Lomonosov Moscow State University, Vorobyovy Gory, 1-2, Moscow, 119991, Russia
}

\date{Accepted XXX. Received YYY; in original form ZZZ}
\pubyear{2018}

\begin{document}
\label{firstpage}
\pagerange{\pageref{firstpage}--\pageref{lastpage}}

{\let\newpage\relax\maketitle}

\begin{abstract}

The present day spectrum of the extragalactic background light (EBL)
in UV, optical and IR wavelengths is the integral result of multiple
astrophysical processes going on throughout the evolution of the
Universe. The relevant processes include star formation, stellar
evolution, light absorption and emission by the cosmic dust. The
properties of these processes are known with uncertainties which
contribute to the EBL spectrum precision. In the present paper we
develop a numerical model of the EBL spectrum while maintaining the
explicit dependence on the astrophysical parameters involved. We
constructed a Markov Chain in the parameter space by using the
likelihood function built with the up-to-date upper and lower bounds
on the EBL intensity. The posterior distributions built with the
Markov Chain Monte Carlo method are used to determine an allowed range
of the individual parameters of the model. Consequently, the star
formation rate multiplication factor is constrained in the range $1.01
< C_{\mbox{sfr}} < 1.69$ at $68\%$ C.L. The method also results in the bounds
on the lifetime, radius, dust particle density and opacity of the
molecular clouds that have large ambiguity otherwise. It is shown that
there is a reasonable agreement between the model and the intensity
bounds while the astrophysical parameters of the best fit model are
close to their estimates from literature.
\end{abstract}

\begin{keywords}
galaxies: star formation -- galaxies: fundamental parameters
\end{keywords}

\section{Introduction}
\label{sec:intro}
Extragalactic background light (EBL) is a radiation emitted by stars
and cosmic dust throughout the whole lifetime of the Universe at
ultraviolet, optical and infrared wavelengths. In ultraviolet and
optical spectral region it is dominated by the star radiation while
the infrared spectrum is determined by emission of the dust. The
understanding of the EBL is an attractive task for several
reasons. First, the origin of EBL is directly connected to the star
formation history and may shed a light on the details of the evolution
of the Universe. Second, EBL is a key element in multiple physical
processes in the intergalactic space. One of these processes is the
attenuation of the high energy gamma-rays by the pair production on
background photons \cite{pair-prod}. Given a photon with the TeV energy, the pair
production cross-section has a maximum for infrared photons, which
constitute an appreciable part of the total flux of EBL. Therefore, an
accurate knowledge of the EBL spectrum is important for reconstruction
of the intrinsic spectra and understanding the emission mechanism of
the distant blazars \cite{Gilmore:2011ks},
\cite{Franceschini:2017iwq}.

Finally, recent studies report anomalies in the reconstructed spectra
of the distant blazars. As was shown 
in~\cite{Horns:2012fx,Rubtsov:2014uga}, the application of the most
conservative EBL models results in the upward breaks in the
intrinsic spectra which are unlikely to be explained by the
source-intrinsic features. The position of the break corresponds to
the energy at which an attenuation becomes significant and the
magnitude of the break is stronger for more distant sources. The
spectra of the blazars will be free from artificial features if the
real density of the EBL flux is at least twice weaker than one in the
most conservative models. The discussion of the possibility of such a
significant reduction required a detailed analysis of the physical
factors accounting for EBL formation. The alternative explanations of
the anomalous transparency of the Universe include new physics with
the hypothetical axion-like particles, see \cite{Troitsky:2016akf}, \cite{DARMA}, \cite{Serpico},
\cite{FRT} for a review and new astrophysics \cite{Kusenko:2010}, \cite{Dzhatdoev:2016ftt}.
In the latter class of models it is assumed that additional 
high-energy gamma-rays are produced as the secondaries in the 
cascades, initiated by the gamma-rays or the ultra-high-energy 
protons emitted at the same sources.

The estimation of the EBL is a complex problem for which several
approaches both experimental and computational have been developed.
First, a large group of methods is based on the observations. A direct
observation of EBL is a challenging experimental problem because of
the dominant contribution of the Zodiacal light and emission of the
Galaxy in the same spectral region. For this reason, these methods in
turn may be split into the two categories. In the first case unwanted
contribution is avoided using the galaxy count method. The method
relies on the deep field observation of the galaxy and the EBL is
estimated as the sum of the fluxes of the discrete
sources~\cite{Keenan:2010na},\cite{Madau:1999yh}. As it follows from
description this approach can only estimate the lower limit of EBL due
to possible undercount of the galaxies beyond the sensitivity level of
the instrument or underestimation of the faintest areas of the objects
detected~\cite{Bernstein:2007}. Another way to make measurement free
of the Zodiacal and Galaxy light corrections is to invoke special
techniques to subtract the background contributions. That results in
EBL considerably higher than that inferred from galaxy count method
\cite{Matsuura:2017lub},\cite{Mattila:2017}.

Another large group of approaches is based on the numerical
calculation of the EBL. These methods differ by the basic assumptions
made and may be divided into several subgroups. The first category is
based on the evolution in time of the galaxy properties. Such an evolution
may be obtained either by direct observations of galaxy evolution or
indirectly estimated according to some prescription.
This method was used in some of the first models of the EBL
\cite{Madau:1998}, \cite{Yichuan:1999}, \cite{Kneiske:2002},
\cite{Kneiske:2004} and it is still relevant due
to the recent large-scale surveys at ultraviolet and infrared
wavelengths \cite{Finke:2009xi}. Another category of approaches uses a
backward evolution of the present day galaxy emissivity according to a
given prescription and accounts for the
change of the star formation rate and other redshift dependent factors
\cite{Franceschini:2017iwq}, \cite{Franceschini:2001yg}, 
\cite{RowanRobinson:2000pr}, \cite{Franceschini:2008tp},
\cite{Dominguez:2010bv}, \cite{Stecker:2016fsg}.

The last scenario for treating EBL is so called the forward evolution
approach. This is the most fundamental and challenging way of
calculation. It starts with the very basic assumptions about the
Universe and it's structure formation. Often the structure formation
process is based on semi-analytic models of cold dark matter merger
trees \cite{Primack:1998wn}, \cite{Gilmore:2011ks}. Being the most
complete, the method allows to pinpoint the physical reasons for
time-dependence of the EBL.

In this Paper we follow the second scenario and present a numerical 
model of the EBL built upon
the dynamics of the star formation and the evolution of stars. Our
goal is to build a model with the explicit dependence on the
underlying astrophysical parameters. The analytic part of the model is
explained in Section~\ref{sec:model} and the numerical part is
presented as a publicly-available
code\footnote{\url{https://github.com/Semk0/EBL-model}}, see Appendix
A. The model allows to explore the parameter space with the Markov
Chain Monte Carlo (MCMC) method. For this purpose in
Section~\ref{sec:method} we build a likelihood function for the EBL
spectrum using the modern upper and lower constrains at multiple
wavelengths. The resulting marginal posterior distributions for the
parameters involved are given in Section~\ref{sec:results}. It is
shown that several parameters, including the multiplication factor of
the star formation rate and the lifetime, radius, dust particle
density and opacity of the molecular clouds of the molecular clouds
are solidly constrained by the EBL observations. The results are
summarized in~\ref{sec:Conclusion}.

\section{Model}
\label{sec:model}
Our goal is to build the empirical model of the formation and
evolution of stars and galaxies. The model we plan to construct will
be then directly translated into the EBL spectrum. It's unavoidable
for the model to be dependent on a number of astrophysical parameters,
which are known with a limited precision. We will keep these
dependencies explicit, so the parameter space may be further probed
with the MCMC technique.

At first the contribution of stars to the EBL will be described, and
then the absorption and re-emission of light by dust will be taken
into account. We use the standard $\Lambda CDM$ cosmological model to
derive the functional dependencies between comoving distance $r$, redshift
$z$ and time $t$.
\begin{equation}
\label{eq:r_z}
	r(z) = \frac{c}{H_0}\int\limits_{0}^{z}\frac{d z}{\sqrt{\Omega_{m}(1+z)^3 + \Omega_{\Lambda}}} 
\end{equation}
\begin{equation}
\label{eq:z_t}
	z(t) = \Big(\frac{\Omega_\Lambda}{\Omega_m}\Big)^{\frac{1}{3}} \big[\sinh(\frac{3}{2}{\Omega_\Lambda}^{\frac{1}{2}} t H_0)\big]^{-\frac{2}{3}} - 1,
\end{equation}
where the Hubble constant $H_0 = 67.8$ km/s/Mpc, the matter density
parameter $\Omega_m = 0.309$, the dark energy density parameter
$\Omega_{\Lambda} = 0.691$ and $c$ is the speed of light. The
parameters of the cosmological model are measured within the high
accuracy \cite{Ade:2015xua} and therefore are considered fixed for
the purposes of the present study. We assume Chabrier initial mass
function (IMF) to describe mass distribution of newborn stars
\cite{Chabrier:2003ki}. Following Gilmore et al. \cite{Gilmore:2011ks}
we assume that the IMF doesn't depend on the redshift:
\begin{equation}
\mathrm{\xi}(m) =
\left\{ \begin{array}{@{\kern2.5pt}lL}
    \hfill \frac{C_{\mbox{\tiny imf}}}{m}e^{-\frac{(\log(m) - \log(m_0))^2}{2D}} & if $m \leqslant 1$,\\
          km^{-a_{\mbox{\tiny imf}}} & if $m > 1$.
\end{array}\right.
\end{equation}
Masses of stars are measured in the units of the Solar mass. The
parameters $m_0$, $D$ and $a$ will undergo variations while $C_{\mbox{\tiny imf}}$ and
$k$ are determined from normalization and continuity. The initial mass
function is bounded form below by the minimal mass $m_{\mbox{\tiny
    min}} = 0.08$. If the mass of the star is less than
$m_{\mbox{\tiny min}}$ then the hydrogen thermonuclear fusion reactions in
the core will not ignite. The upper bound of the mass distribution
$m_{\mbox{\tiny max}}$ is another free parameter of the model. The initial
values for the variable parameters of the IMF are: $m_0 = 0.079$, $D =
0.69$, $a = 2.3$ and $m_{\mbox{\tiny max}} = 100$. For the sake of
convenience we will use mass normalization of the IMF:
\begin{equation}
\int\limits_{m_{\mbox{\tiny min}}}^{m_{\mbox{\tiny max}}} m\xi(m) dm = 1
\end{equation}

The mass of stars, born at redshift $z$ per unit time per unit volume is
described by the star formation rate (SFR) $\psi(z)$. Here we will use
SFR, obtained in \cite{Gilmore:2011ks}. We parametrize the SFR with a
single multiplication factor $C_{\mbox{\tiny sfr}}$ with initial value
equal to unity.

To determine the spectrum of the star with the mass $m$ at the age
$\eta$ near the surface of this star we use stellar evolution tracks
calculated by Hurley et al. \cite{Hurley:2000pk}, which give us radius
$R$, temperature $T$ and lifetime $\eta_{\mbox{\tiny life}}$
dependence on age. Let's define this spectrum as $B(\lambda, m, \eta,
z = 0)$ there the last argument is reserved for the redshift.
\begin{equation}
\label{eq:spec_star_z=0}
B_s(\lambda, m, \eta, 0) = \frac{2 \pi h c^2}{\lambda^5} \frac{1}{e^{\frac{h c}{k T(m, \eta) \lambda}} - 1}
\end{equation}

If we consider the star at the distance $r$ from the detector, the
registered emission will be different from \eqref{eq:spec_star_z=0},
due to geometric dilution and cosmological redshift. The time $\eta$
now will denote the age of the star at the moment of the emission of
radiation registered today. To calculate new spectrum we use a
standard trick: replace $\lambda$ by $\frac{\lambda}{1 + z}$ and
divide the full expression by $(1 + z)^3$. First $(1 + z)$ is
responsible for decreasing the energy of the photons, second is
responsible for decreasing the number of registered photons per unit
time and the last is responsible for broadening of the spectrum.  Also
the expression must be multiplied by the geometrical factor
$\Big(\frac{R(\eta)}{r(z)}\Big)^2$:
\begin{equation}
B_s(\lambda, m, \eta, z) = \Big(\frac{R(m, \eta)}{r(z)}\Big)^2 (1 + z)^{-3} B(\frac{\lambda}{1 + z}, m, \eta, z = 0)
\end{equation}
Distance $r(z)$ is calculated by \eqref{eq:r_z}.

Using the above definitions one may write an expression to determine
the spectrum of a galaxy of age $\eta_{\mbox{\tiny g}}$ and redshift
$z_{\mbox{\tiny g}}$. Moreover, we assume that the star formation
began in all galaxies simultaneously at time $t_i$. The light emitted
at time $t_i$ is now registered with the redshift $z_{\mbox{\tiny
    i}}=z(t_{\mbox{\tiny i}})$. Then the galaxy age $\eta_{\mbox{\tiny
    g}}$ and redshift $z_g$ are related with the expression:
\begin{equation}
\eta_{\mbox{\tiny g}} = t(z_{\mbox{\tiny g}}) - t(z_{\mbox{\tiny i}})
\end{equation}
The function $t(z)$ is the inverse of \eqref{eq:z_t}.
Thus we obtain for galaxy spectrum:
\begin{equation}
G_{\mbox{\tiny s}}(\lambda, z_{\mbox{\tiny g}}) = \int\limits_{m_{\mbox{\tiny min}}}^{m_{\mbox{\tiny max}}} d m \int\limits_{0}^{\eta_{\mbox{\tiny end}}(m)} d \eta' B_{\mbox{\tiny s}}(\lambda, m, \eta', z_{\mbox{\tiny g}})\xi(m) \psi(t(z_{\mbox{\tiny g}}) - \eta')
\end{equation}
Upper limit $\eta_{end}$ in time integral depends on the mass and determined by
\begin{equation}
\label{eq:eta_end}
\eta_{end}(m) = \mbox{min}(\eta_{\mbox{\tiny g}}, \eta_{\mbox{\tiny life}}(m))
\end{equation}
On the other hand the presence of dust and absorbers in the galaxies
and interstellar medium should be taken into account. Following
\cite{Charlot:2000xi} we assume, that the star formation took place
only in the giant molecular clouds which will cover the newborn stars
with the shell of dust and gas. Dust grains consist mostly of graphite
and silicate, so they have non-zero absorption coefficients at visible
and ultraviolet wavelengths. Thus they will be heated by the newborn
stars and re-emit in the infrared part of the spectrum.

Let us now specify the dust model. According to \cite{Charlot:2000xi}
we assume that the birth clouds have finite lifetimes
$\eta_{\mbox{\tiny c}}$. The parameters of the cloud such as particle
number densities and outer radius $R_{\mbox{\tiny c}}$ stay constant
throughout its life, whereas the temperature of different components
may change due to evolution of the stars in the center. Moreover, we
assume for simplicity that dust has constant particle number density
through the entire cloud which is defined by the parameter
$n_{\mbox{\tiny d}}$. We do not account for the other features of the
internal structure of the cloud and following~\cite{Charlot:2000xi}
the optical depth will be treated as a free parameter with the two
characteristics. First one is the normalization, which corresponds to
the optical depth at wavelength $\lambda_0 = 5500 ~\mathring{A}$ and
the second is a spectral slope $n$. Thence, optical depth at the
arbitrary wavelength is determined by
\begin{equation}
\label{eq:tau}
\tau(\lambda) = \tau_{\lambda_0} \Big(\frac{\lambda}{\lambda_0}\Big)^{-n} 
\end{equation}

We assume that the star formation occurs in the center of the birth
cloud, so there exists such a value $\rho \ll R_c$ that all newborn
stars are located in the imaginary sphere $S^{\mbox{\tiny in}}$ with
radius $\rho$. First of all we calculate spectrum $B_c(\lambda, \eta,
\rho)$ of the newborn stars at the age $\eta$ on the boundary of the
sphere $S^{\mbox{\tiny in}}$. On this step we neglect absorption and
dependence on redshift:
\begin{equation}
\begin{aligned}
B_c(\lambda, \eta, \rho) =& \int\limits_{m_{\mbox{\tiny min}}}^{m_{\mbox{\tiny max}}} dm \int\limits_{0}^{\eta_{\mbox{\tiny cend}}(m)} d\eta' \Big(\frac{R(\eta')}{\rho}\Big)^2 \times \\& B(\lambda, m, \eta', z = 0)\xi(m) \psi_c(\eta')
\end{aligned}
\end{equation}
where $\eta_{\mbox{\tiny cend}}(m)$ = min $(\eta_{\mbox{\tiny c}},
\eta_{\mbox{\tiny g}}, \eta_{\mbox{\tiny life}}(m))$ by analogy with
\eqref{eq:eta_end} and $\psi_{\mbox{\tiny c}}(\eta)$ denotes mass of
the matter converted into stars in the cloud at the age $\eta$ per
unit time. The total mass of the gas converted into stars throughout
life of the cloud $M_0$ is given by:
\begin{equation}
\int\limits_{0}^{\eta_c} \psi_c(\eta') d\eta' = M_0
\end{equation}
For the sake of simplicity suppose that $\psi_c(\eta)$ stays constant in time so that:
\begin{equation}
\psi_c = \frac{M_0}{\eta_c}
\end{equation}
On the other hand, $\psi_c$ directly proportional to $\psi(t)$ with
the coefficient $n_{\mbox{\tiny cl}}(t)$ which denotes the number of
clouds per unit volume.
\begin{equation}
\label{cloud_dens}
n_{cl}(t) = \frac{\psi(t)}{\psi_{\mbox{\tiny c}}} = \frac{\psi(t) \eta_{\mbox{\tiny c}}}{M_0}
\end{equation}   
Equation \eqref{cloud_dens} will be used for calculating the spectrum of newborn stars in the galaxy. Applying \eqref{eq:tau} we estimate the absorption for light propagating from $\rho$ to $r$.  
\begin{equation}
B_{\mbox{\tiny c}}(\lambda, \eta, r) = \Big(\frac{\rho}{r}\Big)^2 e^{-\tau_{\lambda_0} (\frac{\lambda}{\lambda_0})^{-n}} B_c(\lambda, \eta, \rho)
\end{equation}
where we again neglected the redshift. Thus $B_c(\lambda, \eta, R_c)$
denotes contribution of the stars to the spectrum of the cloud
calculated on the cloud's boundary. The full spectrum of the cloud
contains an additional contribution of dust. Now we describe the
assumptions regarding properties of dust. Following
\cite{MartinezGalarza:2008jn} the dust grains are spherical with sizes
$a$ distributed by the power law:
\begin{equation}
dn(a, a + da) = C_d a^{-n_{\mbox{\tiny dust}}} da
\end{equation}
Initial value for the slope of distribution $n_{\mbox{\tiny dust}} = 3.5$. As the sizes of grains are constrained, distribution should be truncated with minimal grain size $a_{\mbox{\tiny min}}$ and maximal $a_{\mbox{\tiny max}}$. We will start setting $a_{\mbox{\tiny min}} = 5$ nm and $a_{\mbox{\tiny max}} = 500$ nm. As mentioned above the dust particle number density is constant and allows us to determine normalization constant $C_d$.
\begin{equation}
C_d = \frac{(n_{\mbox{\tiny dust}} - 1)n_d}{a_{\mbox{\tiny min}}^{-n_{\mbox{\tiny dust}} + 1} - a_{\mbox{\tiny max}}^{-n_{\mbox{\tiny dust}} + 1}}
\end{equation}    

Equilibrium temperature $T_d(a, r)$ of dust grain of size $a$ at the distance of $r$ may be found from the energy balance equation:
\begin{equation}
\pi a^2 \int\limits_{0}^{\infty} Q_{\mbox{\tiny abs}}^{a}(\lambda) B_c(\lambda, \eta, r) d\lambda = 4\pi a^2 \int\limits_{0}^{\infty} Q_{\mbox{\tiny abs}}^{a}(\lambda) B_{\mbox{\tiny Pl}}(\lambda, T_d(a, r, \eta))
\end{equation}
where emissivity and absorptivity coefficient of grains with size $a$ are equal due to Kirchoff's law and denoted as $Q_{\mbox{\tiny abs}}^{a}(\lambda)$. $B_{\mbox{\tiny Pl}}(\lambda, T_d(a, r, \eta))$ is the Planck's law of emissivity.

To find out the full contribution of dust we should integrate over the grain size and over the distance $r$. Thus we obtain final formula for calculating dust spectrum:
\begin{equation}
\begin{aligned}
B_d(\lambda, \eta) =& \int\limits_{\rho}^{R_{\mbox{\tiny out}}} dr\ 4\pi r^2 \times\\ &\int\limits_{a_{\mbox{\tiny min}}}^{a_{\mbox{\tiny max}}} da\ \Big(\frac{a}{R_{\mbox{\tiny out}}}\Big)^2 C_d a^{-n_{\mbox{\tiny dust}}} Q_{\mbox{\tiny abs}}^{a}(\lambda) B_{\mbox{\tiny Pl}}(\lambda, T_d(a, r, \eta))
\end{aligned}
\end{equation}

In the mid-infrared the principal contribution comes
  from the polycyclic aromatic hydrocarbons (PAH) molecules, which
  absorb the starlight producing the spectral lines. The corresponding
  cross-sections $\sigma$, damping constants $\gamma$, and cutoff
  energies $\lambda_c$ assuming Lorentzian spectral shape are well
  known and may be found, for example, in \cite{phys-dust}. The PAH spectra is
  calculated as follows:
\begin{equation}
\begin{aligned}
B_{PAH}(\lambda, \eta) =& \int\limits_{\rho}^{R_{\mbox{\tiny out}}}
dr\ 4\pi r^2 \times n_{PAH} \int\limits_{\lambda_c}^{\infty} d\lambda'\ B_c(\lambda', \eta, r) \times \\ 
& \sum_i\ \Big(\sigma_i \frac{\gamma_i  c^3/\lambda^4}{\pi^2 (\frac{c^2}{\lambda^2} - \frac{c^2}{\lambda_0^2})^2 + (\frac{\gamma_i c}{2 \lambda})^2} \Big)
\end{aligned}
\end{equation}
where $n_{PAH}$ is the particle number density of homogeneously distributed PAH molecules, and summing goes over the full set of resonances.

Summarizing the above the full spectrum of the galaxy with redshift $z_g$ is expressed as:
\begin{equation}
G(\lambda, z_g) = G_s(\lambda, z_g) + G_c(\lambda, z_g)
\end{equation}
where:
\begin{equation}
G_s(\lambda, z_g) = \int\limits_{m_{\mbox{\tiny min}}}^{m_{\mbox{\tiny max}}} d m \int\limits_{0}^{\eta_{\mbox{\tiny end}}(m)} d \eta' B_s(\lambda, m, \eta', z_g)\xi(m) \psi(t(z_g) - \eta')
\end{equation}
\begin{equation}
\eta_{\mbox{\tiny end}}(m) = \mbox{Max}(\mbox{Min}(\eta_g, \eta_{\mbox{\tiny life}}(m) - \eta_c), 0)
\end{equation}
\begin{equation}
\begin{aligned}
G_c(\lambda, z_g) = &\int\limits_{0}^{\eta_c} d \eta' (B_d(\lambda, \eta', z_g) + B_{PAH}(\lambda, \eta', z_g) + \\ 
& B_c(\lambda, \eta', R_{out}, z_g))\frac{\eta_c}{M_0}\psi(t(z_g) - \eta') 
\end{aligned}
\end{equation}
Finally, the spectrum of the whole Universe is obtained by integration
over the distance, taking into account Calzetti's attenuation law
$C_{calz}(\lambda)$ for diffuse dust component \cite{Calzetti:1999pg}.
\begin{equation}
U(\lambda) = \int\limits_{0}^{r(z_i)} dr\ 4 \pi r^2 C_{calz}(\lambda) G(\lambda, z_g(r)) 
\end{equation}

\begin{figure*}
  \includegraphics[width=\linewidth]{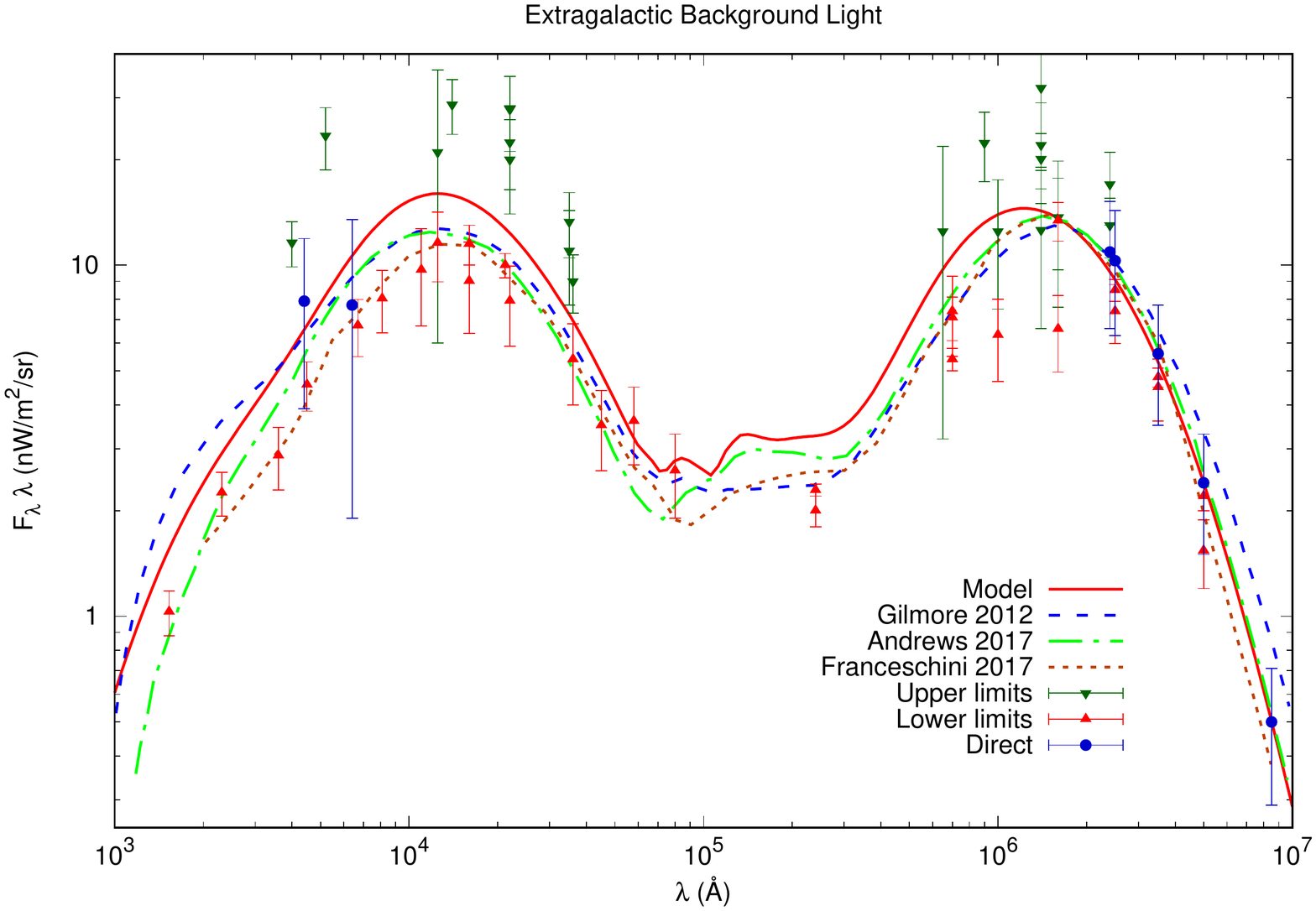}
  \hspace*{\fill}
  \caption{Extragalactic background light spectrum. Red line is our best fit result. Blue line is for Gilmore et al. \protect\cite{Gilmore:2011ks} semi-analytic model. Green and brown lines are for Andrews et al.\protect\cite{Andrews:2017ima} and Franceschini et al.\protect\cite{Franceschini:2017iwq} correspondingly. Upward and downward pointing arrows indicate lower and upper EBL limits, see Table \protect\ref{table:low_cnstr} and Table \protect\ref{tab:up_cnstr}, circles is for direct measurements, Table \protect\ref{tab:dir_cnstr}.
  }
  \label{fig:ebl_spectra}
\end{figure*}

\section{Method}
\label{sec:method}

The model of the Section~\ref{sec:model} keeps an explicit dependence
on a number of parameters. The preferred values and confidence ranges
for the parameters will be calculated using the Markov Chain Monte Carlo
method. The Markov Chain is constructed with the Metropolis algorithm
with the standard likelihood function
\begin{equation}
\mathcal{L} = \displaystyle \prod_i e^{-\frac{(y -
    y_i)^2}{2\sigma_i^2}}\,,
\end{equation}
 where $y_i$ and $\sigma_i$ are the values and errors of the
 experimental points. The one-sided limits are included in the product
 only if $y > y_i$ for lower limits and if $y < y_i$ for upper ones.

The likelihood is based on the three categories of the data points
which are the lower and upper limits of the EBL and direct
measurements. The lower EBL limits usually come from the galaxy count
method and provide strict constraints, see
Table~\ref{table:low_cnstr}. The upper limits appear either from the
direct measurements at the wavelengths where contribution of the
Zodiacal light is significant or from the direct measurements combined
with some special technique to subtract the excess, see
Table~\ref{table:up_cnstr}. We do not include here the constraints
from the gamma-ray observations due to possible uncertainties in AGN's
spectra. For the wavelengths longer than $200 \mu \mbox{m}$ the direct
measurements are used as it is believed that Zodiacal light
contribution is neglectful. We have also added Pioneer 10/11 results
because their measurements were conducted far from the region of the
potential Solar system background affects, see
Table~\ref{table:dir_cnstr}.

\section{Results}
\label{sec:results}

The Markov chain is built as described at
Section~\ref{sec:method}. Each step of the chain requires to repeat
the full calculation of the EBL, so we employ the parallel implementation
of the code with \url{OpenMP}. The chain of the length 42000 is
constructed and made available at the
\footnote{\url{https://github.com/Semk0/EBL-model}}.

Given the model and the chain, we begin with proving the consistency
of the model and then discuss possible systematics. The EBL spectrum
with the best likelihood set of the parameters is shown in Figure
\ref{fig:ebl_spectra}. The parameters inferred from the MCMC are in a
good agreement with the commonly used values, see Table
\ref{table:parameters}. Spectrum demonstrates good accordance in
ultraviolet, optical and far infrared bands with the popular models
\cite{Gilmore:2011ks} and \cite{Franceschini:2008tp}.
\begin{figure*}
  \hfill\includegraphics[width=\linewidth]{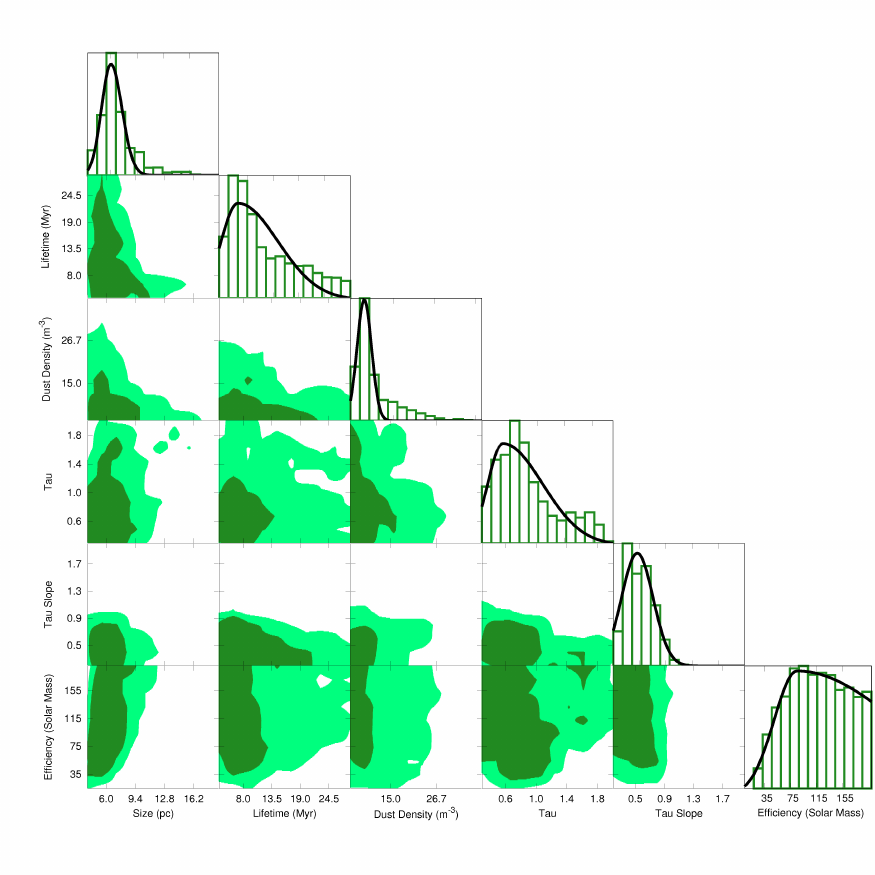}\hspace*{\fill}
  \caption{Distributions of the parameters of the clouds, including radius $R_{\mbox{\tiny c}}$, lifetime $\eta_{\mbox{\tiny c}}$, dust particle density $n_{\mbox{\tiny d}}$, optical depth at $5500 \mathring{A}$ $\tau_{\lambda_0}$, optical depth slope $n$ and cloud efficiency $M_0$.}    
  \label{fig:run4_cloud}
\end{figure*}

We do not account for the AGN's contribution because it is
sub-dominant and moreover is subject to large
uncertainty \cite{Andrews:2017ima}. This contribution may modify the result
considerably for $\lambda < 0.4 \mu \mbox{m}$ and therefore we didn't
account for the experimental points in this range. Another assumption
of the model is that all the cloud's parameters are equal. This means
that all the diversity of the star forming objects is reduced to only
one type of the clouds. We believe this is enough to grasp the main
behaviour of these objects.

Let us now describe the outcomes of the main part of the research. All
the parameters are divided into four group according to the way they
are treated. First group contains the parameters of Chabrier Initial
Mass Function namely $m_{\mbox{\tiny 0}}$, $D$, $a_{\mbox{\tiny
    imf}}$. They are already limited in \cite{Chabrier:2003ki} so we
set the existing confidence intervals as an available freedom for each
parameter. The calculations show that the EBL is insensitive
to $m_{\mbox{\tiny 0}}$ and $D$, while it depends strongly on
$a_{\mbox{\tiny imf}}$. The resulting value is $ a_{\mbox{\tiny imf}} =2.29
^{+0.15}_{-0.24}$ which is more narrow than the limits from
\cite{Chabrier:2003ki}. The parameter $a_{\mbox{\tiny imf}}$ governs
the number of massive stars thus this result indicates that the
massive stars play an important role in the EBL formation.

\begin{figure*}
  \hfill\includegraphics[width=\textwidth]{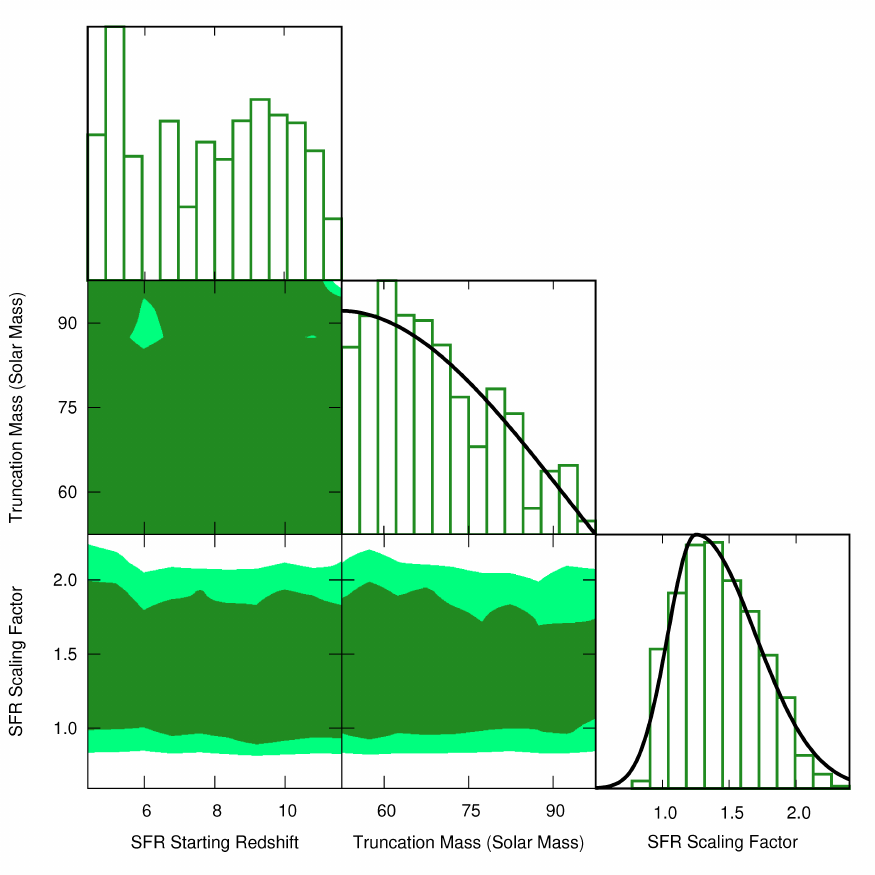}\hspace*{\fill}
  \caption{Distributions of the global parameters of star formation, including redshift at which star formation began $z_{\mbox{\tiny i}}$, IMF truncation mass $m_{\mbox{\tiny max}}$ and SFR scaling factor $C_{\mbox{\tiny sfr}}$}
  \label{fig:run4_glob}
\end{figure*}

The next group contains the dust parameters and includes minimal and
maximal grain sizes and the slope of the distribution $n_{\mbox{\tiny
    dust}}$. We let them vary in a wide range except for lower limit
for the minimal grain sizes and upper limits for the maximal grain
sizes which are $10^{-3} ~\mu$ m and $1 ~\mu$ m correspondingly. After
the simulations we found that the grain sizes are constraint as:
$a_{\mbox{\tiny min}} < 1.75 \cdot 10^{-3}~\mu$ m and $a_{\mbox{\tiny
    max}} < 22.8 \cdot 10^{-3}~\mu$ m. This is due to the higher
emissivity of the bigger grains and thus greater contribution to the
EBL. The slope of the distribution $n_{\mbox{\tiny dust}}$ should be
more than $3.83$. The result may be expected at the large grains
contribute more and their amount should be limited.

The third group collects all the parameters of the clouds, see Figure
\ref{fig:run4_cloud}. Basically, the properties of the molecular
clouds are known from the observations of the Milky Way, for more
details see, for instance, \cite{Murray:2010}. Observations shows that
the radii of the clouds lie in the range from 1 pc to 35 pc with the
largest reach a size of 100 pc. The lifetimes of the clouds with the
mean mass are $17 \pm 4$ Myr \cite{Murray:2010}. The following values
are obtained from the MC for radius $R_{\mbox{\tiny c}} =
6.1^{+1.4}_{-1.2}$ pc and lifetime $\eta_{\mbox{\tiny c}} =
6.0^{+8.5}_{-3.6}$ Myr. These are in a good agreement with the
experimental estimates. The relatively small value of the cloud radius
may be reconciled with the observed larger clouds if we allow the
large clouds to be inhomogeneous and divided into the separate region
of star formation. The dust particle density $n_{\mbox{\tiny d}}$
can't be measured directly but may be derived from the mass of the
cloud. One may derive the star formation efficiency $\epsilon = 0.035$ which
is defined as the ratio between the mass of the newborn stars in the cloud
and the whole mass of the cloud. The calculation uses common assumption that the
gas to dust ratio is 100. The value of $\epsilon$ is in line with the modern
observations~\cite{Murray:2010}.

The last group contains the global parameters of the Universe, see
Figure \ref{fig:run4_glob}. It includes the redshift of the beginning
of the star formation, the mass of the star at which IMF is truncated
and scaling factor for the star formation rate. We obtained that the
EBL does not depend on the epochs with the redshift $4 < z < 10$.
Truncation mass is the maximal mass of
the star in our model. We obtain that the smaller truncation masses
are more probable that the bigger ones. The reasoning for this results
is that the massive stars are too hot and thus overheat the
dust. Finally, the star formation rate scaling factor is well
constrained and at 1-$\sigma$ is $1.25^{+0.44}_{-0.24}$. 

In the calculations above the results of CIBER and ESO
  VLT/FOR were used as the upper limits \ref{table:up_cnstr}. It is
  practical to know how the conclusions of the present paper change if
  these results are interpreted as the direct measurements. For this
  purpose a separate Markov Chain was constructed. The account of
  double-sided CIBER and ESO VLT/FOR gives the higher value of star
  formation rate $C_{\mbox{sfr}} =
  1.76^{+0.14}_{-0.13}$ and lower values of the radius and lifetime 
  of the clouds $R_{\mbox{\tiny c}} < 3.7 \pm 2.0$,
  $\eta_{\mbox{\tiny c}} < 3.1 \pm 2.2$.

\begin{table*}
\centering
\begin{footnotesize}
\caption{Mean values and 1-$\sigma$ C.L. for the MCMC variable parameters in comparison with the value, presented in literature. Dash in MCMC column means that the model does not depend on this parameter.}
\label{table:parameters}
\begin{tabular}{l l l l}
\hline\hline
Parameter name 				& Symbol 					& MCMC 			& literature	 					\\ [0.5ex]
\hline
IMF Parameter ($M_\odot$)	& $m_{\mbox{\tiny 0}}$		& -		  		& $0.079^{+0.021}_{-0.016}$ \cite{Chabrier:2003ki}	\\
IMF Parameter 				& $D$						& -		   		& $0.69^{+0.05}_{-0.01}$ \cite{Chabrier:2003ki}	\\
IMF Parameter 				& $a_{\mbox{\tiny imf}}$	& $2.29 ^{+0.15}_{-0.24}$	& 2.3$\pm$0.3 \cite{Chabrier:2003ki} \\
Dust Particle Minimal Size ($\mu m$)	& $a_{\mbox{\tiny min}}$	& < 1.75 $\cdot 10^{-3}$    & -			\\
Dust Particle Maximal Size ($\mu m$)	& $a_{\mbox{\tiny max}}$	& < 22.8 $\cdot 10^{-3}$	& - 		\\
Dust Distribution Slope 				& $n_{\mbox{\tiny dust}}$	& > 3.83   					& $\sim$3.5 	\\
Radius of the Cloud (pc)				& $R_{\mbox{\tiny c}}$		& $6.1^{+1.4}_{-1.2}$   & 2.5-100 	\cite{Murray:2010}  \\
Lifetime of the Cloud (Myr)				& $\eta_{\mbox{\tiny c}}$	& $6.0^{+8.5}_{-3.6}$  	& $\sim$10 	\cite{Murray:2010}	\\
Dust Particle Density ($m^{-3}$)		& $n_{\mbox{\tiny d}}$	& 6.9$\pm$2.0    			& $\sim$10 	\cite{Murray:2010}	\\
Optical Depth at $5500 ~\mathring{A}$	& $\tau_{\lambda_0}$ 	& $0.59^{+0.57}_{-0.21}$ 	& $\sim$1 	\cite{Charlot:2000xi} \\
Optical Depth Slope 					& $n$       			& 0.47$\pm$0.24  			& $\sim$0.7 \cite{Charlot:2000xi} \\
Cloud Efficiency ($M_\odot$)			& $M_0$     			& $85^{+152}_{-36}$  	 	& - 		\\
Redshift of the beginning of the star formation		& $z_{\mbox{\tiny i}}$	& -   		& $\sim$8 \cite{Gilmore:2011ks}	\\
IMF Truncation Mass ($M_\odot$)			& $m_{\mbox{\tiny max}}$ 	& < 84   				& $\sim$100		\\
SFR scaling factor 						& $C_{\mbox{\tiny sfr}}$	& $1.25^{+0.44}_{-0.24}$& $\sim$1 		\\
\hline

\end{tabular}
\end{footnotesize}
\end{table*}

\section{Conclusion}
\label{sec:Conclusion}
We present the new flexible model of the EBL which is defined as a
function of the astrophysical parameters. It was shown, that the
spectrum based on the common values of these parameters lies in the
experimentally admissible range and is in a good agreement with the
models presented in literature. Further analysis was devoted to study
the spectrum dependence on these parameters and set bounds for each of
them. With the help of the Markov Chain Monte Carlo method we explored
the parameter space and set up constraints on the star formation rate
and parameters of the molecular clouds. Specifically, we have obtained 
the estimate of the IMF slope parameter
$a_{\mbox{\tiny imf}} = 2.29 ^{+0.15}_{-0.24}$. Moreover, assuming the
shape of the SFR from \cite{Gilmore:2011ks} we obtain that the SFR scaling
factor lies in the range from 1.01 to 1.69 at $68\%$~C.L. The latter
implies that the models with the overall decrease of the EBL intensity
are constrained with the observational data.

\begin{table*}
\centering
\begin{footnotesize}
\caption{EBL lower limits used in this paper.}
\label{table:low_cnstr}
\begin{tabular}{c c c}
\hline\hline
$\lambda ~(\mu \mbox{m})$	& Lower Limits~($\mbox{nW}/\mbox{m}^{2}/\mbox{sr}$)	& Experiment	\\ [0.5ex]
\hline
0.153		& 1.03$\pm$0.15				& Galex \cite{Xu:2004zg}					\\
0.231		& 2.25$\pm$0.32				& Galex	\cite{Xu:2004zg}		 			\\ 
0.36		& $2.87^{+0.58}_{-0.42}$	& HDF \cite{Madau:1999yh}					\\ 
0.45		& $4.57^{+0.73}_{-0.47}$	& HDF \cite{Madau:1999yh}					\\ 
0.67		& $6.74^{+1.25}_{-0.94}$	& HDF \cite{Madau:1999yh}					\\ 
0.81		& $8.04^{+1.62}_{-0.92}$	& HDF \cite{Madau:1999yh}					\\ 
1.1			& $9.71^{+3.0}_{-1.9}$		& HDF \cite{Madau:1999yh}					\\ 
1.6			& $9.02^{+2.62}_{-1.68}$	& HDF \cite{Madau:1999yh}					\\ 
2.2			& $7.92^{+2.04}_{-1.21}$	& HDF \cite{Madau:1999yh}					\\ 
1.25		& 11.7$\pm$2.6				& Subaru \cite{Keenan:2010na}					\\ 
1.6			& 11.5$\pm$1.5				& Subaru \cite{Keenan:2010na}					\\ 
2.12		& 10.0$\pm$0.8				& Subaru \cite{Keenan:2010na}					\\
3.6			& 5.4$\pm$1.4				& Spitzer/IRAC \cite{Fazio:2004kx} 	 			\\ 
4.5			& 3.5$\pm$0.9				& Spitzer/IRAC \cite{Fazio:2004kx}				\\ 
5.8			& 3.6$\pm$0.9				& Spitzer/IRAC \cite{Fazio:2004kx}		\\ 
8.0			& 2.6$\pm$0.7				& Spitzer/IRAC \cite{Fazio:2004kx}		\\ 
24			& 2.29$\pm$0.09				& Spitzer/MIPS \cite{Bethermin:2010} 	\\ 
70			& 5.4$\pm$0.4				& Spitzer/MIPS \cite{Bethermin:2010} 	\\ 
70			& 7.4$\pm$1.9				& Spitzer/MIPS \cite{Frayer:2006qq}		\\ 
24			& 2.0$\pm$0.2				& Spitzer/MIPS \cite{Chary:2004ht}		\\ 
70			& 7.1$\pm$1.0				& Spitzer/MIPS \cite{Dole:2006de}		\\ 
160			& 13.4$\pm$1.7				& Spitzer/MIPS \cite{Dole:2006de}			\\ 
100			& 6.33$\pm$1.67				& Herschel/PACS	\cite{Berta:2010}			\\ 
160			& 6.58$\pm$1.62				& Herschel/PACS	\cite{Berta:2010}			\\ 
250			& 8.5$\pm$0.6				& BLAST	\cite{Devlin:2009}				\\ 
350			& 4.8$\pm$0.3				& BLAST \cite{Devlin:2009}				\\ 
500			& 2.2$\pm$0.2				& BLAST \cite{Devlin:2009}				\\ 
250			& 7.4$\pm$1.42				& Herschel/SPIRE \cite{Bethermin:2012}	\\ 
350			& 4.5$\pm$0.9				& Herschel/SPIRE \cite{Bethermin:2012}	\\ 
500			& 1.54$\pm$0.34				& Herschel/SPIRE \cite{Bethermin:2012}	\\[1ex]
\hline
\end{tabular}
\end{footnotesize}
\end{table*}

\begin{table*}
\centering
\begin{footnotesize}
\caption{EBL upper limits used in this paper.}
\label{table:up_cnstr}
\begin{tabular}{c c c}
\hline\hline
$\lambda ~(\mu \mbox{m})$ 	& Upper Limits~($\mbox{nW}/\mbox{m}^{2}/\mbox{sr}$)	& Experiment \\ [0.5ex]
\hline
0.4		& 11.6$\pm$1.7		& ESO VLT/FORSE \cite{Mattila:2017} \\
0.52	& 23.4$\pm$4.7		& ESO VLT/FORSE \cite{Mattila:2017} \\
1.25	& 21$\pm$15			& COBE/DIRBE \cite{Levenson:2007} \\
2.2		& 20$\pm$6			& COBE/DIRBE \cite{Levenson:2007} \\
3.5		& 13.3$\pm$2.8		& COBE/DIRBE \cite{Levenson:2007} \\
1.25 	& 54$\pm$16.8		& COBE/DIRBE \cite{Cambresy:2001ei}	\\
2.2		& 27.8$\pm$6.7		& COBE/DIRBE \cite{Cambresy:2001ei} \\
1.4 	& $28.7^{+5.1}_{-3.3}$  & CIBER \cite{Matsuura:2017lub} \\
3.6		& $9.0^{+1.7}_{-0.9}$	& COBE/DIRBE \cite{Levenson:2008jc} \\
2.2		& 22.4$\pm$6.0		& COBE/DIRBE \cite{Gorjian:1999zi} \\
3.5		& 11.0$\pm$3.3		& COBE/DIRBE \cite{Gorjian:1999zi} \\
65		& 12.5$\pm$9.3		& Akari \cite{Matsuura:2011}	\\
90		& 22.3$\pm$5.0		& Akari \cite{Matsuura:2011}	\\
140		& 20.1$\pm$3.6		& Akari \cite{Matsuura:2011}	\\
160		& 13.7$\pm$4.0		& Akari \cite{Matsuura:2011}	\\
140		& 12.6$\pm$6.0		& COBE/FIRAS \cite{Fixsen:1998kq}   \\
160		& 13.7$\pm$6.1		& COBE/FIRAS \cite{Fixsen:1998kq}   \\
140		& 32$\pm$13			& COBE/DIRBE \cite{Schlegel:1997yv} \\
240		& 17$\pm$4			& COBE/DIRBE \cite{Schlegel:1997yv} \\
100		& 12.5$\pm$5.0		& COBE/DIRBE \cite{Wright:2003tp} \\
140		& 22$\pm$7			& COBE/DIRBE \cite{Wright:2003tp} \\
240		& 13.0$\pm$2.5		& COBE/DIRBE \cite{Wright:2003tp} \\ [1ex]
\hline
\end{tabular}
\end{footnotesize}
\end{table*}

\begin{table*}
\centering
\begin{footnotesize}
\caption{Direct measurements of the EBL used in this paper.}
\label{table:dir_cnstr}
\begin{tabular}{c c c}
\hline\hline
$\lambda ~(\mu \mbox{m})$ 	& Direct Measurements~($\mbox{nW}/\mbox{m}^{2}/\mbox{sr}$)	& Experiment \\ [0.5ex]
\hline
0.44	& 7.9$\pm$4.0 		& Pioneer 10/11	\cite{Matsuoka:2011} 	\\
0.64 	& 7.7$\pm$5.8	 	& Pioneer 10/11 \cite{Matsuoka:2011} 	\\
240		& 10.9$\pm$4.3 		& COBE/FIRAS \cite{Fixsen:1998kq} 		\\
250 	& 10.3$\pm$4.0 		& COBE/FIRAS \cite{Fixsen:1998kq} 		\\
350		& 5.6$\pm$2.1 		& COBE/FIRAS \cite{Fixsen:1998kq} 		\\
500 	& 2.4$\pm$0.9 		& COBE/FIRAS \cite{Fixsen:1998kq} 		\\
850 	& 0.5$\pm$0.21 		& COBE/FIRAS \cite{Fixsen:1998kq}		\\ [1ex]

\hline
\end{tabular}
\end{footnotesize}
\end{table*}

\section*{Acknowledgements}
The authors are indebted to Maxim Pshirkov, Dmitry Semikoz,
Sergey Troitsky and Valery Rubakov for inspiring discussions. This
work has been supported by the Russian Science Foundation grant
14-22-00161. The numerical part of the work is performed at the
cluster of the Theoretical Division of INR RAS. A.K.
is supported by the fellowship of Basis Foundation.

\section*{References}

\bibliographystyle{mnras}
\bibliography{mybib} 

\section*{Appendix A}
\label{sec:Appendix_A}
The numerical implementation of model from Section~\ref{sec:model} is
available at \url{https://github.com/Semk0/EBL-model}. There are two
possible scenarios of using the code. First, the EBL spectrum may be
constructed with the help of \url{spec.cpp} module. The parameters of
the EBL are defined in the \url{in/param.txt} file. One may fine the
full description of the parameters in the \url{spec.cpp} file. The
second usage scenario is the optimization of the parameters for the
specific EBL constraints. It is assumed that the data point are
separated into the three categories: upper limits, lower limits and
direct measurements. The program read these points from the following
files \url{in/upper_limits.txt}, \url{in/lower_limits.txt},
\url{in/direct.txt}. The make command compiles the code and starts
optimization.

\bsp	
\label{lastpage}
\end{document}